\documentstyle[iau,twoside]{article}
\begin{document}

\renewcommand{\baselinestretch}{1.05}


\section[Spectroscopy of Ultra-Cool Dwarfs]{Spectroscopy of Ultra-Cool Dwarfs}

Ultra-cool dwarfs have been defined as dwarfs with spectral types of M7 or later
(Kirkpatrick, Henry, \& Irwin 1997) and thus include the new L and T spectral 
classes. The previous IAU report on ultra-cool dwarfs in 1999 described the 
optical and IR spectral features that characterize these spectral types. 
In this report, I review the latest progress on the spectral classification
of ultra-cool dwarfs, which are divided into two categories:
old dwarfs in the field ($>1$~Gyr) and their young progenitors in the nearest 
star-forming regions and open clusters (1-100~Myr). 
Because the former are nearby and the latter are young, these two groups are
relatively bright and lend themselves to discovery and detailed study.
I conclude by discussing some of the major results from two recent meetings.
Technically, the luminosity class of young ultra-cool 
objects is closer to subgiant than dwarf; 
nevertheless, I include them in this report on ultra-cool dwarfs.

\subsection[The Field]{The Field}

Three years ago, L-type objects had been discovered in abundance ($\sim100$),
while only a dozen T-type sources were known.
Since then, 2MASS, SDSS, and other surveys have identified another
150 L dwarfs and 20 T dwarfs free-floating in the field
(Cuby et al. 1999; Strauss et al. 1999; Kirkpatrick et al. 2000; 
Reid et al. 2000; Tsvetanov et al. 2000; Leggett et al. 2000;
Burgasser et al. 1999, 2000b, 2002a; Hawley et al. 2002; Liu et al. 2002b).
The first L and T dwarfs were discovered as companions
(GD165B, Becklin \& Zuckerman 1988; Gl229B, Nakajima et al. 1995), and
companion searches have continued to reveal new members of these spectral
classes. Ultra-cool dwarfs orbiting field stars have been found
at wide angular separations through 2MASS images and followup spectroscopy by
Kirkpatrick et al. (2001), Wilson et al. (2001), Gizis, Kirkpatrick, \&
Wilson (2001), and Burgasser et al. (2000a). Companions that are likely
to be ultra-cool have appeared at small angular separations in images
obtained with HST and ground-based telescopes using adaptive optics (AO)
(Mart{\'\i}n et al. 2000b; Reid et al. 2001b; Els et al. 2001; Close
et al. 2002a, 2002b), some of which have been observed with newly available
AO spectroscopy (Potter et al. 2002; Goto et al. 2002; Liu et al. 2002a).

Spectral classification of ultra-cool dwarfs began with optical data
for M and L types (Kirkpatrick et al. 1999; Mart{\'\i}n et al. 1999b). 
Classification at these types has since been developed at
IR wavelengths (Reid et al. 2001a; Leggett et al. 2001; Testi et al. 2001).
The numerous discoveries of ultra-cool dwarfs have produced a well-sampled
spectral sequence into late L and T types, enabling the definition of spectral
subclasses in the T regime (Burgasser et al. 2002a; Geballe et al. 2002).
To first order, the T spectral class was originally defined by the presence of
methane absorption at IR wavelengths. However, under the accepted classification
schemes for L and T types, late L dwarfs exhibit methane absorption at 
3.3~$\mu$m (Noll et al. 2000), and possibly at 2.2~$\mu$m as well 
(McLean et al. 2001; Nakajima, Tsuji, \& Yanagisawa 2001).
The temperatures below which methane absorption appears in each photometric
band have been compared to theoretical predictions by Schweitzer et al. (2002).
In addition to temperature, clouds are probably responsible for much of the
variation in spectral features among L and T types (Burgasser et al. 2002b).

\subsection[Open Clusters and Star-forming Regions]{Open Clusters and 
Star-forming Regions}

Searches for ultra-cool dwarfs in nearby open clusters and star-forming regions
have been motivated by the fact that brown dwarfs 
($M\mathrel{\hbox{\rlap{\hbox{\lower4pt\hbox{$\sim$}}}\hbox{$<$}}}0.075$~$M_\odot$)
are brightest and warmest when they are young, which enables their detection 
down to very low masses. For instance, objects near the hydrogen burning mass
limit have spectral types of M6-M7 when they are younger than 100~Myr, and
eventually cool to mid-L types after several Gyr. In other words, a given
ultra-cool spectral type corresponds to a much lower mass (by an order of
magnitude) in a star-forming region than in the field.

The Pleiades open cluster (125~Myr) was the first site to prove fruitful
in surveys for young ultra-cool dwarfs (Stauffer, Hamilton, \& Probst 1994;
Mart{\'\i}n et al. 2000a). Members of this cluster have been found with spectral
types as late as early L ($\sim0.035$~$M_\odot$, Mart{\'\i}n et al. 1998). 
In the last few years, the emphasis has shifted to very young associations
and star-forming clusters ($<10$~Myr), where the brown dwarfs should be even
more luminous.
Free-floating ultra-cool objects have been identified with optical and IR
spectroscopy toward IC~348 (Luhman 1999), Chamaeleon~I 
(Neuh\"{a}user \& Comer\'{o}n 1999; Comer\'{o}n, Neuh\"{a}user, \& Kaas 2000), 
Ophiuchus (Luhman, Liebert, \& Rieke 1997; Wilking, Greene, \& Meyer 1999; 
Cushing, Tokunaga, \& Kobayashi 2000), 
Taurus (Mart{\'\i}n et al. 2001a), the Orion Nebula 
Cluster (Hillenbrand 1997; Lucas et al. 2001), $\sigma$~Ori (B\'{e}jar et 
al.\ 1999; Zapatero Osorio et al. 2000; Barrado et al. 2001), and the 
TW~Hya association (Gizis 2002).
In addition, companion searches have located late-type secondaries to young
stars in the Taurus star-forming region (White et al. 1999) and in the 
associations of Tucanae (Lowrance et al. 2000; Guenther et al. 2001) and 
TW~Hya (Lowrance et al. 1999; Neuh\"{a}user et al. 2000).
Most of these sources are late-M (0.01-0.1~$M_\odot$), while a smaller 
number of L-type objects (0.005-0.01~$M_\odot$) have been found toward
Orion and $\sigma$~Ori.

I now describe the spectral features that vary between ultra-cool 
dwarfs in the field and their young counterparts in clusters.
Over time, the processes of convection 
and nuclear burning deplete Li at the surfaces of young low-mass stars
and brown dwarfs. This depletion occurs faster for more massive objects, 
resulting in a boundary between non-Li and Li objects that evolves
to lower masses, and thus fainter luminosities and cooler temperatures.
For instance, the Li depletion boundary is near the hydrogen burning mass limit 
for an age of $\sim100$~Myr.
Stauffer, Schultz, \& Kirkpatrick (1998), Barrado et al. (1999), and
Stauffer et al. (1999) have used this phenomenon to estimate ages for the 
Pleiades (125~Myr), IC~2391 (53~Myr), and $\alpha$~Per (90~Myr) open clusters. 
In star-forming clusters, ultra-cool sources are too young to have depleted
their Li and thus exhibit strong absorption at 6707~\AA\
(Mart{\'\i}n, Basri, \& Zapatero Osorio 1999a). At M types, the presence of 
Li absorption can be used to distinguish young late-type members of a cluster 
from foreground field dwarfs. However, the time scale for Li depletion 
becomes longer with lower masses, so that some L-type members of the field 
($>1$~Gyr) can exhibit Li absorption. As a result, at L types Li is not a 
definitive indicator of membership in a young cluster.


Because ultra-cool objects are extremely faint at optical wavelengths,
the high-resolution data necessary for Li measurements are difficult to obtain.
Instead, youth and membership are more easily determined with the 
gravity-sensitive Na~I and K~I absorption lines, which are stronger than Li 
and reside at brighter regions of the optical spectrum. 
At late M types, these features are weaker in members of open clusters
and star-forming regions than in field dwarfs, which has been attributed to 
the lower surface gravities of the former (Mart{\'\i}n, Rebolo, \& Zapatero 
Osorio 1996; Luhman et al. 1998). 
However, at types later than $\sim$L0, these lines are difficult to use as
diagnostics of youth because they become weaker and the optical flux continues 
to decrease.
Ultra-cool field dwarfs and young sources can be differentiated 
at IR wavelengths, which is where these objects are most easily observed.
Steam absorption at each end of the H and K bands results in a plateau in 
the spectra of dwarfs, while for young sources this plateau is less apparent 
and the spectrum is instead more sharply peaked (Lucas et al. 2001).



L-type objects found toward star-forming regions are sometimes referred to 
as ``planetary-mass objects" (Lucas et al. 2001; Zapatero Osorio et al. 2000)
because L0 corresponds roughly to the deuterium burning mass limit 
(0.013-0.015~$M_\odot$) for ages of $<10$~Myr 
(Burrows et al. 1997; Baraffe et al. 1998).
The relationship between the L types reported by Lucas et al. (2001) from IR 
spectra and the standard optically-based classifications for field dwarfs is
unclear, but these sources are likely to be young members of Orion because they
show IR steam features that are distinctive from those of field dwarfs, and
in a manner that is consistent with predicted variations with surface gravity
(Allard et al. 2001).
However, many of the L-type objects toward the $\sigma$~Ori cluster
lack convincing spectroscopic evidence of youth and membership in the cluster.
Although the early L source found by Zapatero Osorio et al. (1999) appears
to have Na~I and K~I line strengths that are indicative of youth,
the remaining objects are too faint ($I>21$) for such measurements.
Barrado y Navascu\'es et al. (2001) 
reported strong H$\alpha$ emission for some these sources, which would be
evidence for youth, but most of these detections appear to have marginal 
significance by the spectra presented in that study.
Indeed, the distinctive shape of the steam absorption bands observed for young 
objects in Orion by Lucas et al. (2001) is not present in the $\sigma$~Ori 
objects (Mart{\'\i}n et al. 2001b), which may indicate that the latter are 
foreground field dwarfs rather than cluster members. 





For the spectral classification of young objects at late-M types, Luhman 
(1999) found that the various spectral features between 6000 and 9000~\AA\ 
are best matched with averages of spectra for standard dwarfs and giants. 
Because some of the IR steam bands become stronger at a fixed temperature
from dwarfs to young objects (Allard et al. 2001), optically-classified young 
sources are probably the appropriate standards when using IR spectra to
measure spectral types for young late-type objects.


\begin{references}
\reference Allard, F., Hauschildt, P. H., Alexander, D. R., Tamanai, A.,
\& Schweitzer, A. 2001, \apj, 556, 357
\reference Baraffe, I., Chabrier, G., Allard, F., \& Hauschildt, P. H. 1998,
\aap, 337, 403
\reference Barrado y Navascu\'es, D., Stauffer, J. R., Song, I., \& Caillault,
J.-P. 1999, \apj, 522, L53
\reference Barrado y Navascu\'es, D., Zapatero Osorio, M. R., B\'ejar, V. J. 
S., Rebolo, R., Mart{\'\i}n, E. L., Mundt, R., \& Bailer-Jones, C. A. L. 2001,
\aap, 377, L9
\reference Becklin, E. E., \& Zuckerman, B. 1988, Nature, 336, 126
\reference B\'ejar, V. J. S., Zapatero Osorio, M. R., \& Rebolo, R. 1999,
\apj, 521, 671
\reference Burgasser, A. J., et al. 1999, \apj, 522, L65
\reference Burgasser, A. J., et al. 2000a, \apj, 531, L57
\reference Burgasser, A. J., et al. 2000b, \aj, 120, 1100
\reference Burgasser, A. J., et al. 2002a, \apj, 564, 421
\reference Burgasser, A. J., et al. 2002b, \apj, 571, 151
\reference Burrows, A., et al. 1997, \apj, 491, 856
\reference Close, L. M., Potter, D., Brandner, W., Lloyd-Hart, M., Liebert,
J., Burrows, A., \& Siegler, N. 2002a, \apj, 566, 1095
\reference Close, L. M., Siegler, N., Potter, D., Brandner, W., \& Liebert, J.
2002b, \apj, 567, L53
\reference Comer\'{o}n, F., Neuh\"{a}user, R., \& Kaas, A. A. 2000, \aap, 359,
269
\reference Cuby, J. G., Saracco, P., Moorwood, A. F. M., D'Odorico, S.,
Lidman, C., Comer\'on, F., \& Spyromilio, J. 1999, \aap, 349, 41
\reference Cushing, M. C., Tokunaga, A. T., \& Kobayashi, N. 2000, \aj, 119, 
3019
\reference Els, S.G., Sterzik, M.F., Marchis, F., Pantin, E., Endl, M., \&
K\"urster, M. 2001, \aap, 370, L1
\reference Geballe, T. R., et al. 2002, \apj, 564, 466
\reference Gizis, J. E. 2002, \apj, 575, 484
\reference Gizis, J. E., Kirkpatrick, J. D., \& Wilson, J. C. 2001, \aj, 121,
2185
\reference Goto, M., et al. 2002, \apj, 567, L59
\reference Guenther, E. W., Neuh\"{a}user, R., Hu\'elamo, N., Brandner, W.,
\& Alves, J. 2001, \aap, 365, 514
\reference Hawley, S. L., et al. 2002, \aj, 123, 3409
\reference Hillenbrand, L. A. 1997, \aj, 113, 1733
\reference Kirkpatrick, J. D., et al. 1999, \apj, 519, 802
\reference Kirkpatrick, J. D., et al. 2000, \aj, 120, 447
\reference Kirkpatrick, J. D., Dahn, C. C., Monet, D. G.; Reid, I. N.,
Gizis, J. E., Liebert, J., \& Burgasser, A. J. 2001, \aj, 121, 3235
\reference Kirkpatrick, J. D., Henry, T. J., \& Irwin, M. J. 1997, \aj, 113,
1421
\reference Leggett, S. K., et al. 2000, \apj, 536, L35
\reference Leggett, S. K., Allard, F., Geballe, T. R., Hauschildt, P. H., \&
Schweitzer, A. 2001, \apj, 548, 908
\reference Liu, M. C., Fischer, D. A., Graham, J. R., Lloyd, J. P., Marcy,
G. W., \& Butler, R. P. 2002a, \apj, 568, 107
\reference Liu, M. C., Wainscoat, R., Mart{\'\i}, E. L., Barris, B., \& Tonry,
J. 2002b, \apj, 568, L107
\reference Lowrance, P. J., et al. 1999, \apj, 512, L69
\reference Lowrance, P. J., et al. 2000, \apj, 541, 390
\reference Lucas, P. W., Roche, P. F., Allard, F., \& Hauschildt, P. H. 2001,
\mnras, 326, 695
\reference Luhman, K. L. 1999, \apj, 525, 466
\reference Luhman, K. L., Brice\~{n}o, C., Rieke, G. H., \& Hartmann, L. W. 
1998, \apj, 493, 909
\reference Luhman, K. L., Liebert, J., \& Rieke, G. H. 1997, \apj, 489, L165
\reference Mart{\'\i}n, E. L., et al. 1998, \apj, 507, L41
\reference Mart{\'\i}n, E. L., et al. 2000a, \apj, 543, 299
\reference Mart{\'\i}n, E. L., Basri, G., Zapatero Osorio, M. R. 1999a, \aj, 
118, 1005
\reference Mart{\'\i}n, E. L., Delfosse, X., Basri, G., Goldman, B., Forveille,
T., \& Zapatero Osorio, M. R. 1999b, \aj, 118, 2466
\reference Mart{\'\i}n, E. L., Dougados, C., Magnier, E., M\'{e}nard, F.,
Magazz\`{u}, A., Cuilandre, J.-C., \& Delfosse, X. 2001a, \apj, 561, L195
\reference Mart{\'\i}n, E. L., Koresko, C. D., Kulkarni, S. R., Lane, B. F.,
\& Wizinowich, P. L. 2000b, \apj, 529, L37
\reference Mart{\'\i}n, E. L., Rebolo, R., \& Zapatero Osorio, M. R. 1996, 
\apj, 469, 706
\reference Mart{\'\i}n, E. L., Zapatero Osorio, M. R., Barrado y Navascu\'es, 
D., B\'ejar, V. J. S., \& Rebolo, R. 2001b, \apj, 558, L117
\reference McLean, I. S., Prato, L., Kim, S. S., Wilcox, M. K., Kirkpatrick,
J. D., \& Burgasser, A. 2001, \apj, 561, L115
\reference Nakajima, T., Oppenheimer, B. R., Kulkarni, S. R., Golimowski, D.
A., Matthews, K., \& Durrance, S. T. 1995, Nature, 378, 463
\reference Nakajima, T., Tsuji, T., \& Yanagisawa, K. 2001, \apj, 561, 119
\reference Neuh\"{a}user, R., \& Comer\'{o}n, F. 1999, \aap, 350, 612
\reference Neuh\"{a}user, R., Guenther, E. W., Petr, M. G., Brandner, W.,
Hu\'elamo, N., \& Alves, J. 2000, \aap, 360, 39
\reference Noll, K. S., Geballe, T. R., Leggett, S. K., \& Marley, M. S. 2000,
\apj, 541, 75
\reference Potter, D., Martin, E. L., Cushing, M. C., Baudoz, P., Brandner, W.,
Guyon, O., \& Neuh\"{a}user, R. 2002, \apj, 567, L133
\reference Reid, I. N., Burgasser, A. J., Cruz, K. L., Kirkpatrick, J. D., \&
Gizis, J. E. 2001a, \aj, 121, 1710
\reference Reid, I. N., Gizis, J. E., Kirkpatrick, J. D., \& Koerner, D. W.
2001b, \aj, 121, 489
\reference Reid, I. N., et al. 2000, \aj, 119, 369
\reference Schweitzer, A., Gizis, J. E., Hauschildt, P. H., Allard, F.,
Howard, E. M., \& Kirkpatrick, J. D. 2002, \apj, 566, 435
\reference Stauffer, J. R., et al. 1999, \apj, 527, 219
\reference Stauffer, J. R., Hamilton, D., \& Probst, R. G. 1994, \aj, 108, 155
\reference Stauffer, J. R., Schultz, G., \& Kirkpatrick, J. D. 1998, \apj,
499, 199
\reference Strauss, M. A., et al. 1999, \apj, 522, L61
\reference Testi, L., et al. 2001, \apj, 552, L147
\reference Tsvetanov, Z. I., et al. 2000, \apj, 531, L61
\reference White, R. J., Ghez, A. M., Reid, I. N., \& Schultz, G. 1999, \apj,
520, 811
\reference Wilking, B. A., Greene, T. P., \& Meyer, M. R. 1999, \aj, 117, 469
\reference Wilson, J. C., Kirkpatrick, J. D., Gizis, J. E., Skrutskie, M. F.,
Monet, D. G., \& Houck, J. R. 2001, \aj, 122, 1989
\reference Zapatero Osorio, M. R., B\'ejar, V. J. S., Mart{\'\i}n, E. L.,
Rebolo, R., Barrado y Navascu\'es, D., Bailer-Jones, C. A. L., \& Mundt, R.
2000, Science, 290, 103
\reference Zapatero Osorio, M. R., B\'ejar, V. J. S., Rebolo, R.,
Mart{\'\i}n, E. L., \& Basri, G. 1999, \apj, 524, L115
\end{references}
\end{document}